\documentclass[useAMS,usenatbib]{mn2e}

\title[Is the high mass binary pulsar PSR J 1614-2230 a latent magnetar?]{Is the high mass binary pulsar PSR J 1614-2230 a latent magnetar?}
\author[ Vikram Soni]{Vikram Soni$^{1}$\thanks{E-mail:
 v.soni@airtelmail.in (VS)}
\footnotemark[1]\thanks{}
\\
$^{1}$ Centre for Theoretical Physics, Jamia Millia University, New Delhi 110025, India \\}

\begin{document}

\date{   }

\pagerange{\pageref{firstpage}--\pageref{lastpage}} \pubyear{}

\maketitle

\label{firstpage}

\begin{abstract}
We consider the newly found high mass and low magnetic field binary
pulsar PSR J1614-2230 in  a model in which magnetars owe their
strong magnetic fields to a high baryon density,
 magnetized core.
In our magnetar model all neutron stars above a certain threhold
mass are magnetars. This confronts us with the very basic paradox as
to why this high mass star, PSR J1614-2230, remains a pulsar and
shows no magnetar characteristics. This is a star that has acquired
its large mass by accretion from its binary companion over 5
gigayears.

In this work we show that the maximum rate of energy gain from the
strong interaction phase transition from this slow accretion does
not allow for high enough interior temperature for ambipolar
transport of the magnetic field to the surface of the star and thus
the PSR J 1614-2230 remains latent and does not become an emergent
magnetar.
\end{abstract}


\section{Introduction}

Generally, magnetars are neutron stars with surface magnetic fields
( $10^{14(15)}$ G) a thousand times larger than that of an average
pulsar. The magnetars have spin down ages of $10^{3}-10^{5}$ years.
Over this period, they emit a quasisteady radiative luminosity of
$10^{34}$ - $10^{36}$ erg/s. Some of them emit repeated flares or
bursts of energy typically of $10^{41} - 10^{44}$ erg. The periods
of magnetars  fall in a surprisingly narrow window of 5-12 s (see
(\citep{b17}) for a review). For conventional magnetars with large
periods, the energy emitted in both quiescent emission and flares
far exceeds the loss in their rotational energy. The most likely
energy source for these emissions is their magnetic energy
(\citep{b8,b20}), yet there is no evidence of a decrease in their
surface magnetic fields with time \citep{b21}.

There have been many attempts to explain some of this physics of
which the most popular is the magnetar model of Duncan and Campbell
(\citep{b8,b20}), which is otherwise known as the {\em dynamo
mechanism for magnetars}. This model requires the collapse of a
large mass progenitor to a star which starts life with a period
close to a millisecond. Such a fast rotation can amplify the
inherited pulsar valued field of $10^{12}$ G to $10^{15}$ G.
However, as pointed out in earlier works (\citep{b3,b12}), several
observations on magnetars are hard to understand from such a  model.

These works were based on a model which argues that magnetars have
larger masses  than pulsars and that their higher density cores
undergo a strong interaction phase transition to a magnetized ground
state. In these works (\citep{b3,b12}) it was shown that it may be
possible to explain many unusual features of magnetars if they are
born with a highly magnetized core created by the strong
interaction. Initially  the core magnetic field is
 screened by the surrounding plasma of electrons, protons and neutrons.As the screening
 currents dissipate the core field is transported first from the core
 to the crust and then from the crust to the surface (\citep{b12}) powering the enhanced X-ray
flux and enhancing the surface magnetic field .

In this magnetar model all neutron stars above a certain threshold mass are magnetars. This confronts us with
the puzzle of why the binary pulsar, PSR J1614-2230, whose mass hass been
recently determined (\citep{b7}) to be $\simeq 2 $ solar masses, remains a pulsar and shows no magnetar
 characteristics.
 In this work we show that the rate of energy gain from the accretion induced strong interaction phase
transition does not allow for high enough interior temperature for ambipolar transport of the magnetic field
and thus the PSR J 1614-2230 remains latent.

\section{The Model}
Pulsars, which have radii of $ \simeq 10 km $ and surface magnetic
fields of $ \simeq 10^{10}- 10^{12}$ G, are believed to inherit such
fields from their progenitors, which are stars of radius $ \simeq
10^{6} km $ and  magnetic fields of $ \simeq 1 - 10^{2}$ G  due to
'conservation' of magnetic flux during stellar collapse
(\citep{b23}).

Our starting point, to make the distinction between pulsars and
magnetars, is that pulsars exist up to some threhold mass $ M_T$ and
central density . For larger masses and consequently higher central
density, the core of neutron stars undergo a phase transition giving
rise to magnetars. When the core density exceeds about three times
the nuclear density new and interesting phases may appear. Of these,
there are some that can naturally produce a very large magnetization
of the core . One possible such state is  a neutral pion condensate
groundstate (\citep{b6,b2,b19}), while there are others
(\citep{b14,b10})
 that are independent of the presence of a pion condensate .
These ground states can produce very large core magnetic fields
of the order of $10^{16} $ G.

As the core magnetization grows in time, the magnetic field of the
core will initially be screened by the highly conducting exterior
plasma in accordance with Lenz's law. Eventually the Lenz
(screening) currents dissipate establishing the full dipolar field
due to core magnetization outside the core . In the process of this
dissipation, energy is carried away from the star as thermal
effects, neutrino emission etc. A central feature of our model is
that there is a characterstic time during which {\em ambipolar
diffusion} (\citep{b9}) carries the core field to the crust with
copious neutrino emission. The emerging magnetic field then cleaves
the crust, increasing resistivity and the shielding currents get
dissipated to power the radiative emissions from magnetars.

\section{The new ~2 solar mass binary neutron star }

In this work we consider the recently reported (\citep{b7})discovery
of highest mass neutron star, the binary pulsar PSR J1614-2230,
using the precision technique of Shapiro delay (\cite{b7}). It has a
low field , $B =  1.8 \cdot 10^{8} $ G, no enhanced X-ray flux and
no flares and a  spin down age of  $ 5 \cdot 10^{9} $ years,
associated with recycled pulsars. Additionally, PSR J1614-2230, has
an orbital period of 8.7 days with the final mass of the donor (
white dwarf) being $\simeq 0.5  M_s $. In our model, for all neutron
stars above a certain threhold mass, magnetar charcteristics emerge
when a high density magnetized core is created at birth by the
strong interaction. However, this is a star that has acquired its
large mass by accretion from its binary companion, which also spins
it up to a period of 3.15 milliseconds, over 5 gigayears. We will
demonstrate here that this is the reason that, PSR J1614-2230, does
not show magnetar features.

\section{ Equation of state, fast rotation and the maximum mass of the star}

The spun up binary, PSR J1614-2230, ( $M \simeq 2 M_s $) has a period of
3.15 milliseconds.  A neutron star with such rapid rotation will have an
 enhanced maximum mass, due the strong centrifugal forces that  push out
 matter in the star and counteract gravity.

For non relativistic fermions, like neutrons, the fermi pressure
goes as density to the five  third power and can effectively
counteract gravity to form stable stars. For relativistic quark
 matter ground  states the fermi pressure goes as density to the
four thirds, which is not strong enough to hold off gravity and
 results in an  instability that sets the maximum mass of the star.
Stars with quark matter cores have yet another problem;
they have a stiff non relativistic neutron exterior  pushing in a
softer relativistic quark interior - an unstable situation. In this
case a star with a quark core is stable only if the nuclear matter to
quark matter transition takes place in a small window at low pressure (\cite{b19}).

 It is also for this reason that most neutron stars with quark matter cores
and in particular with meson condensates have smaller maximum
allowed masses. In the absence of rotation,the maximum mass of
neutron stars with a quark matter core normally works out to be
around  $M_{max} \simeq 1.6 M_s $ as recorded in the compilation of
Lattimer and Prakash  (\citep {b15}) and observed by Demorest et al
(\citep {b7}). This was also confirmed in the results of (\citep
{b19}).

Cook et al (\citep {b5,b4}) have looked at the allowed masses of
rotating neutron stars using mass shed and radial instability
limits. With a fast rotation that corresponds to a period of a few
milliseconds, the maximum mass of a star with a soft equation of
state EOS ( for example, a quark matter core), could be raised to
$M_{max} \simeq 1.8 M_s $,  still falling well short of PSR
J1614-2230, ( $M \simeq 2 M_s $).

The existence of such a large mass neutron star would then eliminate
all typical soft equations of state associated with quark matter (
with/without comdensate) interiors.

In contrast, the maximum mass of a purely nuclear star governed by
the APR equation of state of Akmal et al (\citep{b1})  is $ 2.2 M_s
$. If we factor in rotation this mass will be even  higher. It may
then not be unreasonable to expect that a star governed by APR
equation of state (EOS), even with a pion condensate, could have a
mass of 2 solar masses.

For details on the nuclear equation of state,with a $ \pi_0 $
condensate we refer the reader to previous work and references
therein (\citep{b6,b2,b1,b19}). Other possibilities for creating
magnetized cores without pion condensation have been considered by
Kutschera and Wojcik (\citep{b14}) and by Haensel and Bonazzola
(\citep{b10}). These works provide a different scenario  for
creating a core using conventional nuclear physics
 (fermi liquid theory) that is independent of pion condensation .

We assume that PSR J1614-2230 is a purely nuclear star with a magnetized core. We will
now turn to our model stars which have magnetic cores that are created by a high density
phase transition .

\section {Magnetar by birth or accretion}

Neutron stars with a large mass could result either,(i) from
the core collapse of a rather massive star or (ii) by heavy accretion
onto a neutron star in a binary system. In either case,
if the final mass exceeds $M_T$, a magnetic  core will form.
Will one expect to see a magnetar in all such situations? The
answer depends on the details of the thermal structure in the
neutron star interior.

(i) A newly born neutron star in a stellar collapse has a very
hot interior, facilitating ambipolar diffusion and allowing the
strong field to emerge at the surface in a short time.

The time scale of ambipolar diffusion to transport the magnetic
field to the crust for a neutron, proton, electron plasma in the
interior of a neutron star, have been worked out by Goldreich and
Reisenegger (\citep{b9}). Their estimates show that ambipolar
diffusion has a dissipation time scale of

 $t _{ap} \simeq   10^4 \cdot B_{16}^{-2} \cdot T_{8.5}^{-6}$ years,

where $ B_{16}$ is the local magnetic field strength in units of
$10^{16}$ G and $T_{8.5}$ is the temperature in units of $10^{8.5}$
K, a typical value in the interior of a young neutron star.The more
massive the neutron star, the larger will be the size of the
magnetic core and the quicker will the strong field emerge to the
surface.

(ii) A neutron star accumulating matter via accretion, on
the other hand, is old and its interior is relatively cold. Heating
due to the accretion process is not expected to raise
interior temperature above  $10^{7.5 - 8} $ K (\citep{b24}). As we show in what follows,
the extreme temperature sensitivity of the ambipolar diffusion rate will then
delay the emergence of the field at the surface, perhaps for
such a long time that the magnetar property would never be
visible. This may be the reason why the surface magnetic field PSR J1614-2230
 remains low (\citep{b7}) despite its mass growing to a large value.

\section{ Features of an accreted nuclear matter magnetar with a magnetized  core}

1) To begin with, let us deal with the question if this binary star,
PSR J1614-2230, was born a magnetar. In the context of our model
this means, did it have a large enough mass at birth to have a
magnetized core.

If it was born with a magnetized core its surface magnetic field
would be large ,  $ B > 10^{13} $ gauss, and like the other
magnetars, would have emerged at the surface in   $\simeq 10^{5-6} $
years, which is not the case. We can conclude that PSR J1614-2230
was not born a magnetar.

2) As pointed out by van den Heuvel (\citep {b22}) if an accreted
neutron star was born with a typical pulsar mass $\simeq 1.4  M_s $
then its short pulse (millseconds period) could be the result of a
long lasting mass accretion of at least $\simeq 0.1  M_s $. However,
if this was the case for PSR J1614-2230, it would imply an
abnormally large mass accretion,  $\simeq 0.6  M_s $ of mass. The
larger the mass accreted the larger the spin up - in this case the
star should have spun up to a period less than a millisecond.

Van den Heuvel (\citep {b22}) also points to another scenario that
can arise from a large mass progenitor, $M > 19 M_s$.  In  this
scenario, which he considers more likely for PSR J1614-2230, the
star can be born with a mass larger than $\simeq 1.7  M_s $. In this
case the accreted mass would be of the order of $ 0.2 - 0.3$ solar
mass, which is more reasonable.  Independently, Lin et al
(\citep{b16}) have carried out extensive simulations of an accreted
neutron star which acquires its mass from high mass X ray binary .
Given the parameters of PSR J1614-2230, an orbital period of 8.7
days and the final mass of the donor ( white dwarf) of $\simeq 0.5
M_s $, they find that this can only happen if the initial neutron
star mass is of the same order, $\simeq 1.7  M_s $.

It is then probable that PSR J1614-2230 is to be understood as a
purely nuclear star with an initial
 mass of at least, $M  \simeq 1.7 M_s$. If it starts life below the threshold mass of a magnetar,
it follows that the threshold mass for a ( nuclear) magnetar is $M_T
> 1.7 M_s$. With an observed mass of almost 2 solar masses, it
follows that PSR J1614-2230 can pick up at most,  $ 0.2 - 0.3$
solar mass by accretion from its companion.

3) The total mass energy added to the star must then be less than
 $  0.3 \cdot M_s   \simeq  0.6 \cdot 10^{33} $ gm

Since the star was not born a magnetar, only a fraction of this
accreted mass would go into making the magnetized core. We note that
conventional (born not accreted) magnetars, with a substantial core
of radius, $R_c \simeq 2-3 km$ and an average density of $ 10^{15} $
gm/cc, would have a core of mass $M_c \simeq 0.03 M_s$. For PSR
J1614-2230, we must take account of the fact that accretion is
accompanied by spin up to millisecond periods which reduces the
nucleon  density (\citep {b11}) in the core. It is then likely that
the high density ($ \simeq 10^{15} $ gm/cc) core mass have an upper
limit of $ M_{core} \simeq  0.03 \cdot M_s   \simeq  0.06 \cdot
10^{33} $ gm.

Taking a typical energy release in the strong phase transition (eg.
to a pion condensed core) of $ \simeq 10 $ Mev/nucleon $ = 1.5 \cdot
10^ {-5}$ erg/nucleon (\citep{b6,b2,b1,b19}), the total energy
release from the core works out to, $ \simeq 10^{51} $ ergs ( upper
limit).



 4) Now, accretion keeps adding mass to the core as the star builds up in a time scale set by its age.  In contrast to born magnetars the strong interaction phase transition happens gradually over the age of the star and so the energy release thereof.  Given the elecctromagnetic opacity of the surrounding plasma it will heat up the rim of the core
 and allow for neutrino emission. It would well
be that less than one perecent of the energy released will be converted (\citep{b13,b24}) to heat. In this
case  from  a total $\simeq 10^{51} $ ergs of energy we may have a leftover  balance of $ \simeq 10^{49} $ ergs
. Of this balance a part will be used in generating the core magnetic field and the shielding currents that
screen it.

5) For a magnetised core of $2-3$ km and a magnetic field of  $ B \simeq 10^ {16} $ gauss the magnetic energy
contained in the core is

 $ E _M ^{core} \simeq 10^{48} $ erg

A similar but somewhat larger amount of energy sits in the screening
currents that shield the core. Note that the magnetic energy is of
the same order as the balance energy left after accounting for the
loss from neutrino emission . These screening currents dissipate
during the age of the star. An estimate of the average energy flux
is given by dividing the total screening current energy release by
the age of the star -  $  5 $ Gigayears.

$ {\dot E_s} \simeq 10^{30(31)}$ ergs /sec

6)  It is good to keep in mind that for a conventional magnetar with
a similar core a similar energy release happens  rather quickly via
the strong interaction as the core formation is completed shortly
after the star is born. This energy heats up the interior of the
star and a large fraction may be emitted as neutrinos. Also, an
amount of energy of order, $ E _M ^{core} \simeq 10^{48(49)} $ erg,
goes into creating the core magnetic field and the consequent
shielding. However,  now the shielding currents get dissipated in $
\simeq 10^5 $ years. This yields  an average energy flux of

  $ {\dot E_s} \simeq  10^{35(36)}$ ergs /sec

which can give rise to  (\citep{b13}) interior temperatures of $
10^{8.5}  $ K, required for efficient ambipolar diffusion in
conventional magnetars.

This is at least four orders of magnitude larger than the energy
flux from PSR J1614-2230. It would appear that the energy flux of
from PSR J1614-2230 may not be able to sustain  interior
temperatures of $  10^{8.5}  $ K.

7)  Let us next look at the internal temperatures of regular accreted, spun up neutron stars without any
additional heating sources in the interior( without any magnetic cores and dissipation of sheilding currents).
 The authors of reference (\citep{b24}) have looked at interior temperatures of accretion based spun up
neutron stars with and without pion condensed cores to find that the accreted neutron stars with pion
condensed interiors cool faster and have interior temperatures of  less than $  10^{7.5}  $ K, whereas stars
 with normal n,p,e interiors can have slightly higher interior temperatures of up to $\simeq 10^{8} $  K.

Such temperatures for spun up accretion pulsars are also indicated
by the work of Potekhin et al (\citep {b18}). They find that pulsars
with accreted material envelopes are different from normal pulsars
with iron (Fe) envelopes - the former having smaller interior
temperatures than the latter as evidenced in their Fig 1 (\citep
{b18}). For regular accreted pulsars they indicate surface
temperatures in the range of $ T \simeq 10^{5-6} $  K \citep
{b18}and interior temperatures of $ T_I \simeq 10^{7.5} $ K .

\section{The time scale of ambipolar diffusion to transport the magnetic field to the crust}
The dissipation time scale of ambipolar diffusion to transport the
magnetic field to the crust for a neutron, proton, electron plasma,
in the interior of a neutron star, have been estimated by Goldreich
and Reisenegger (\citep{b9}),

 $t _{ap} \simeq   10^4 \cdot B_{16}^{-2} \cdot T_{8.5}^{-6}$ years,

This star has a very large core magnetic field and yet does not
manifest as a magnetar. The reason for this has to then be the
internal temperature not reaching a high enough value for ambipolar
diffusion to be effective. Recall, that from the above formula, the
time of transport by ambipolar diffusion goes inversely as the sixth
power of the temperature .

For an accreting star, with a magnetised core ,we have magnetic
fields of $10^{16}$ G  at the core surface and fields of
$10^{14(15)}$ G at the surface of the star. If we take the mean
field in the interior of the star to be $10^{15} $ G and mean
interior temperature to be, $\simeq 10^{7.5}$ K , then the  the
ambipolar diffusion formula gives a typical travel time of $ \simeq
10^{12}$  years to reach the surface. This is larger than the age of
the star.   In our model this completes the understanding of why the
magnetar core fields for spun up, accreted magnetars like, PSR
J1614-2230,  are not manifest.

\section{Acknowledgements}
Firstly, I would like to acknowledge the contribution of  N. D.
Haridass who was closely involved in a  substantial part of this
work. I would like to thank the University Grants Commission, ICTP
Trieste, and Centre for Theoretical Physics, Jamia Millia for
support during this work.

\label{lastpage}
\end{document}